\documentclass[runningheads]{llncs}
\usepackage{graphicx}
\usepackage{booktabs}
\usepackage{multirow}
\usepackage{siunitx}
\usepackage{xcolor}
\usepackage{cite}
\usepackage{amssymb,amsmath,amsfonts,multirow,rotate}

\makeatletter
\newcommand{\labitem}[2]{%
\def\@itemlabel{\textbf{#1}}
\item
\def\@currentlabel{#1}\label{#2}}
\makeatother

\begin{document}
\title{Multilayer Network Analysis: The Identification of Key Actors in a Sicilian Mafia Operation}
\titlerunning{Key actors in a Sicilian Mafia operation}
%
\author{Annamaria Ficara\inst{1,2}\orcidID{0000-0001-9517-4131} \and
Giacomo Fiumara\inst{2}\orcidID{0000-0003-1528-7203} \and
Pasquale De Meo\inst{3}\orcidID{0000-0001-7421-216X} \and
Salvatore Catanese\inst{2}\orcidID{0000-0002-0369-8235}}
\authorrunning{Ficara et al.}
%
\institute{University of Palermo, DMI Department, via Archirafi 34, 90123 Palermo, Italy \\
\email{aficara@unime.it} \\
\and University of Messina, MIFT Department, V.le F. S. D’Alcontres 31, 98166 Messina, Italy \\
\email{\{gfiumara,scatanese\}@unime.it}\\
\and University of Messina, DICAM Department, Viale G. Palatuci 13, 98168 Messina, Italy\\
\email{pdemeo@unime.it}}
\maketitle              
\begin{abstract}
Recently, Social Network Analysis studies have led to an improvement and to a generalization of existing tools to networks with multiple subsystems and layers of connectivity. These kind of networks are usually called multilayer networks. Multilayer networks in which each layer shares at least one node with some other layer in the network are called multiplex networks. Being a multiplex network does not require all nodes to exist on every layer. In this paper, we built a criminal multiplex network which concerns an anti-mafia operation called ``Montagna'' and it is based on the examination of a pre-trial detention order issued on March 14, 2007 by the judge for preliminary investigations of the Court of Messina (Sicily). ``Montagna'' focus on two Mafia families called ``Mistretta'' and ``Batanesi'' who infiltrated several economic activities including the public works in the north-eastern part of Sicily, through a cartel of entrepreneurs close to the Sicilian Mafia. Originally we derived two single-layer networks, the former capturing meetings between suspected individuals and the latter recording phone calls. But some networked systems can be better modeled by multilayer structures where the individual nodes develop relationships in multiple layers. For this reason we built a two-layer network from the single-layer ones. These two layers share 47 nodes. We followed three different approaches to measure the importance of nodes in multilayer networks using degree as descriptor. Our analysis can aid in the identification of key players in criminal networks.

\end{abstract}

\section{Introduction}

During his lifetime, each individual continuously deal with multiple social networks. He does it with no effort and this does not mean that it is a trivial activity which can be overlooked.
The connections between people through multiple types of relational ties represent only one possible view of a problem already known long before the field of social network analysis (SNA) was developed.
Looking only at a single type of relational tie within a single social network risks either defining a world where different kinds of relationships are ontologically equivalent or overlooking the invisible relationships emerging from the interactions among different types of ties.
For a long time, these interactions have largely been studied within a single-layer perspective and one of the most effective SNA tools to measure social interactions has been the simple graph. A simple graph is defined as a set of nodes, also called actors (\textit{i.e.} individuals or organizations) with edges between them, also called links or connections (\textit{i.e.} relational ties such as friendship relationships) and with no edges connecting a node to itself~\cite{Dickison2016}. 
\hfill\break\indent 
According to Wasserman and Faust~\cite{wasserman1994social}, social networks contain at least three different dimensions: a structural dimension corresponding to the social graph (\textit{e.g.} actors and their relationships); a compositional dimension describing the actors (\textit{e.g.} their personal information); and an affiliation dimension (\textit{e.g.} members of the same family or organization). These three dimensions provide a minimal description needed to understand the full complexity of social structures. 
An alternative conceptual approach to dealing with the same set of problems is to think of multiple relationships as a set of connected levels, or layers, forming a single multidimensional social network~\cite{Dickison2016}.
In fact, a social network with nodes and/or edges can be organized into multiple layers, where each layer represents a different kind of node or edge, a different social context, a different community, a different online social network (OSN), and so on.
The analysis of multiple layers can provide knowledge that is not present in each layer when layers are considered independently of each other.
\hfill\break\indent
Kivel\"a et al.~\cite{Kivela2014} review and discuss many of the relevant works on the topic. Then they try to unify the literature by introducing a general framework for multilayer networks~\cite{DeDomenico2013, BOCCALETTI2014, DeDomenico2015, Catanese2017}. Such framework can represent the different notions of networks (\textit{e.g.} single-layer or monoplex networks~\cite{DeDomenico2013, Degani2016}, multiplex networks~\cite{Battiston2014, Ribalta2014, Nicosia2015}, interdependent networks, networks of networks) by simply introducing cumulative constraints on the general model~\cite{Tomasini2015}.
\hfill\break\indent 
Multilayer social networks appear in a number of different contexts, where data are characterized by different sizes, different natures (\textit{e.g.} online, offline, hybrid), and different layer semantics (\textit{e.g.} contact, communication, time, context, etc.). 
Many multirelational networks, that is actors connected by multiple types of ties, have been collected during SNA studies. These networks are often characterized by a small size, because they were often collected through offline questionnaires or interviews and they can be very useful in qualitatively checking the behavior and results of new methods~\cite{Dickison2016}. 
\hfill\break\indent
An interesting multirelational network about criminal relationships is described by Bright et al.~\cite{Bright2015} who focused on eight types of edges related to the exchange of a particular resource (\textit{e.g.} drugs, money) in a criminal network with $128$ actors.
\hfill\break\indent 
We propose a new real multilayer criminal network derived from two simple graphs described in our earlier works~\cite{Ficara2020, Calderoni2020, Cavallaro2020, Cavallaro2021}, whose datasets are publicly available on Zenodo~\cite{Zenodo2020}. These simple graphs capture the physical meetings and phone calls among suspects in an anti-mafia operation called ``Montagna'', concluded in 2007 by the Public Prosecutor's Office of Messina (Sicily). Starting from these two simple graphs, we created an undirected and weighted multilayer network with two layers called Meetings and Phone Calls, $154$ nodes and $439$ edges. 
\hfill\break\indent 
The Sicilian Mafia~\cite{Gambetta1996, Paoli2004, Paoli2008} is one of the most renowned criminal organisations (\textit{i.e.} clans, families, gangs, syndicates) whose social structure analysis generated great scientific interest~\cite{Kleemans2008}. 
SNA has become an important tool for the study of criminal networks~\cite{Ferrara2014, Ferrara2014a, Agreste2016} and it can be used to describe the structure and functioning of a criminal organisation, to construct crime prevention systems~\cite{Berlusconi2016, Calderoni2020}, to identify leaders within a criminal organisation~\cite{Johnsen2018} or to evaluate police interventions aimed at dismantling and disrupting criminal networks~\cite{Duijn2014, Villani2019, Cavallaro2020}.
\hfill\break\indent 
To identify leaders within a criminal network, we have to use a family of measures aimed at identifying the most important actors in a social network \cite{wasserman1994social}.
The family of centrality measures is probably the most widely applied set of SNA tools in practical contexts. Centrality~\cite{ficara2020correlations} is an intrinsically relational concept, because to be central, an actor needs to have relations \cite{Dickison2016}. An actor might be important because he is connected to a large number of different nodes or because he is connected to other important nodes. An actor can also be considered important because his absence would result in a loosely connected social network made of many isolated components. Centrality is also often described in terms of the power that an actor could receive from it (\textit{e.g.} an actor strategically located within a network will have a high control over the information flowing through the network).
\hfill\break\indent 
In this paper, we focus on the importance of nodes in multilayer networks. We used three different approaches (see details in Sect.~\ref{sec:results}) to compute the nodes' degree (\textit{i.e.} the number of edges adjacent to it) and to identify the $20$ most important nodes (\textit{i.e.} the key actors) in our criminal multilayer network.

\section{Background}
\label{sec:background}

\subsection{Single-layer networks}
\label{subsec:single}

A \textit{single-layer network} is a simple graph~\cite{wasserman1994social} which can be defined as a tuple $G = (V,E)$ where $V = \{v_1,v_2,...,v_N\}$ is the set composed by $N$ nodes and $E = \{l_1,l_2,...,l_L\}$, $E \subseteq V \times V$, is the set of edges, whose generic element $l_k$ represents the edge existing between a pair of nodes $(v_i,v_j)$.
The graph edges sometimes have weights, which indicate the strength (or some other attribute) of each connection between the nodes.

From a purely mathematical point of view, the information contained in both sets $V$ and $E$ can be represented in an $A$ matrix of dimension $N \times N$ called the \textit{adjacency matrix}~\cite{wasserman1994social}, whose generic element $a_{ij}$ is defined as
\begin{equation}
a_{ij} = \begin{cases}
$1$ &\text{if $v_i \rightarrow v_j \qquad i,j = 1,2,...,N$}\\
$0$ &\text{otherwise}
\end{cases}
\end{equation}
In undirected graphs, $a_{ij} = a_{ji}$ for all $i \neq j$ and therefore the adjacency matrix $A$ will be symmetrical: $A = A^T$ (where $T$ is the transpose representation).

It is possible to define some descriptive measures able to highlight particular characteristics of a network. For example, we define the {\it degree} of a node $d(v_i)$ as the number of nodes adjacent to it~\cite{wasserman1994social} or as the cardinality of the set of neighbors of that node:
\begin{equation}
d(v_i) = |\mathcal{N}(v_i)| = |\{v_j : \exists (v_i,v_j) \lor \exists (v_j,v_i), j \neq i \}|    
\end{equation}

The degree assumes a discrete value between a minimum of $0$ when a node is isolated (\textit{i.e.} it is not connected to any other node) and a maximum of $N - 1$ if the node is connected to all other nodes in the network.
The degree of a node can be calculated by adding the columns (or rows) of the adjacency matrix $A$~\cite{wasserman1994social}:
\begin{equation}
d(v_i) = \sum_{j=1}^{N}a_{ij} = \sum_{i=1}^{N}a_{ij} 
\end{equation}

The concepts presented so far can be expressed using an alternative notation, which makes use of tensors and Einstein's notation~\cite{DeDomenico2013, Degani2016}.

Given the canonical basis in the vector space $\mathbb{R}^N$, $\xi = \{e_1, e_2,...,e_N\}$ where $e_i = (0,...,0,1,0,...,0)^T$ is $1$ in the \textit{i}th component, and $0$ otherwise. Given a set of $N$ nodes $v_i$ (where $i = 1, 2,...,N$ and $N \in \mathbb{N}$), we associate with each node a \textit{state} that is represented by the canonical vector $e_i$ in the vector space $\mathbb{R}^N$. 
A node $v_i$ can be related with each other and the presence and the intensity of such relationships in the vector space is indicated using the tensor product~\cite{Abraham1988} (\textit{i.e.} the \textit{Kronecker product}) $\mathbb{R}^N \otimes \mathbb{R}^N = \mathbb{R}^{N \times N}$. Thus, second-order (\textit{i.e.} rank-2) canonical tensors are defined by $E_{ij} = e_i \otimes e^T_j$ (where $i, j = 1, 2,...,N$). 
The \textit{relationship tensor} can consequently be written as
\begin{equation}
W = \sum_{i=1}^{N}\sum_{j=1}^{N} w_{ij}E_{ij} = \sum_{i=1}^{N}\sum_{j=1}^{N} w_{ij} e_i \otimes e^T_j
\end{equation}
where the intensity of the relationship from node $v_i$ to node $v_j$ is indicated by $w_{ij}$. 

The matrix $W$ is an example of an \textit{adjacency tensor} and, in the context of single-layer networks, it is just a $N \times N$ matrix of a weighted graph with $N$ nodes which is equivalent to the adjacency matrix $A$.

The covariant notation by Ricci and Levi-Civita~\cite{Ricci1900} can be used to write an adjacency tensor. In this case, a row vector $a \in \mathbb{R}^N$ can be represented using its covariant and controvariant components which are the vector $a_\alpha$ (\textit{i.e.} $\alpha = 1,2,...,N$) and its dual vector $a^\alpha$ (\textit{i.e.} a column vector in the Euclidean space). A linear combination of tensors in the canonical basis can be used to represent the adjacency tensor $W$:
\begin{equation}
W^\alpha_\beta = \sum_{i=1}^{N}\sum_{j=1}^{N} w_{ij} e^\alpha(i) e_\beta(j) = \sum_{i=1}^{N}\sum_{j=1}^{N} w_{ij} E^\alpha_\beta(ij)
\label{adjtensor}
\end{equation}
where $e^\alpha(i)$ and $e_\beta(j)$ are respectively the $\alpha$th and the $\beta$th components of the $i$th contravariant and $j$th covariant canonical vectors in $\mathbb{R}^N$, $E^\alpha_\beta(ij) \in \mathbb{R}^{N \times N}$ is the tensor in the canonical basis which represents $E_{ij}$ (\textit{i.e.} the tensor product of the canonical vectors assigned to nodes $v_i$ and $v_j$).

Define the 1-vector $u^\alpha = (1, 1,...,1)^T \in \mathbb{R}^N$ and let $U_\alpha^\beta = u_\alpha u^\beta$ be the second-order tensor whose elements are all equal to 1 (\textit{i.e.} a so-called \textit{1-tensor}). The \textit{degree vector} is calculated adding up all the columns of the adjacency tensor defined in Eq.~\ref{adjtensor}:
\begin{equation}
k_\beta = W^\alpha_\beta u_\alpha
\label{degreesingle}
\end{equation}
It is possible to calculate the degree of node $v_i$ by projecting the degree vector onto the $i$th canonical vector: 
\begin{equation}
k(i) = k_\beta e^\beta (i)
\end{equation}

\subsection{Multilayer networks}
\label{subsec:multi}

Kivel\"a et al.~\cite{Kivela2014} define a multilayer network as the most general structure which can be used to represent any kind of network. At the base of this structure, there is the elementary concept of graph, defined in Subsect.~\ref{subsec:single}.
The representation of networks at multiple levels or with multiple types of edges (or with other similar features) requires structures that have \textit{layers} in addition to nodes and edges.
Moreover, the concept of \textit{aspect} can be defined as a feature of a layer representing one dimension of the layer structure (\textit{e.g.} the type of an edge or the time at which an edge is present)~\cite{Tomasini2015}. 
More specifically, an ``elementary layer'' is an element of one of the possible sets of layers from a specific aspect and the term ``layer'' refers to a combination of elementary layers from all aspects. 

A \textit{multilayer network} can be defined as a quadruplet $M = (V_M , E_M , V , L)$. 
$V_M \subseteq V \times L_1 \times \cdot \cdot \cdot \times L_d$ is the set of the node-layer combinations, that is the set of layers in which a node $v_i \in V$ is present. $E_M \subseteq V_M \times V_M$ is the edge set containing the set of pairs of possible combinations of nodes and elementary layers. $V$ is the set of all nodes independently from the layer. $L = \{L_a\}^d_a=1$ is the sequence of sets of elementary layers such that there is one set of elementary layers $L_a$ for each aspect $a$. If $d = 0$, the multilayer network $M$ reduces to a single-layer network. If $d = 1$, then $M$ reduces to a multiplex network.

Using multiple layers, it is possible to represent different types of edges: those among nodes in the same layer, called \textit{intralayer edges}, and those among nodes in different layers, called \textit{interlayer edges}. For this reason, the concepts of intralayer and interlayer adiacency tensor are introduced~\cite{DeDomenico2013}. The \textit{intralayer adiacency tensor} $C^\alpha_\beta (\widetilde{k}\widetilde{k})$ is defined as the relationships among nodes in the same layer $\widetilde{k}$ which is indicated by the second-order tensor $W^\alpha_\beta (\widetilde{k})$, where $\alpha, \beta = 1, 2, ..., N$ as defined in Eq.~\ref{adjtensor}.
Indices which refer to layers are distinguished from those which correspond to nodes using the tilde symbol.
The second-order \textit{interlayer adjacency tensor} $C^\alpha_\beta (\widetilde{h}\widetilde{k})$ is introduced instead to encode information about relationships between nodes on different layers (\textit{e.g.} a node $v_i$ from layer $\widetilde{h}$ can be connected to a node $v_j$ in an other layer $\widetilde{k}$). The interlayer adjacency tensor $C^\alpha_\beta (\widetilde{h}\widetilde{k})$ corresponds to the intralayer adjacency tensor $W^\alpha_\beta (\widetilde{k})$ when the same layer $\widetilde{k}$ is represented by a couple of layers.

Following a similar approach to the one used to define the adjacency tensor for single-layer networks (see Eq.~\ref{adjtensor}), the vector $e^{\widetilde{\gamma}}(\widetilde{k})$ (where $\widetilde{\gamma},\widetilde{k} = 1,2,...,L$) of the canonical basis in the space $\mathbb{R}^L$ is introduced. In this definition, the vector components are indicated by a greek index while the $k$th canonical vector is indicated by a latin index. Therefore the second-order tensors, which correspond to the canonical basis in the space $\mathbb{R}^{L \times L}$, are constructed as
\begin{equation}
E^{\widetilde{\gamma}}_{\widetilde{\delta}} (\widetilde{h}\widetilde{k}) = e^{\widetilde{\gamma}}(\widetilde{h}) e_{\widetilde{\delta}}(\widetilde{k})
\label{secordtens}
\end{equation}
The \textit{multilayer adjacency tensor}~\cite{Degani2016} can be written from Eq.~\ref{secordtens}, using a tensor product between the adjacency tensors $C^{\widetilde{\alpha}}_{\widetilde{\beta}} (\widetilde{h}\widetilde{k})$ and the canonical tensors
$E^{\widetilde{\gamma}}_{\widetilde{\delta}} (\widetilde{h}\widetilde{k})$. A fourth-order tensor is obtained as 
\begin{equation}
M^{\alpha\widetilde{\gamma}}_{\beta\widetilde{\delta}} = \sum_{\widetilde{h}=1}^{L}\sum_{\widetilde{k}=1}^{L} C^\alpha_\beta (\widetilde{h}\widetilde{k}) E^{\widetilde{\gamma}}_{\widetilde{\delta}} (\widetilde{h}\widetilde{k})
\label{muladjtens}
\end{equation}
The second-order interlayer adjacency tensor $C^\alpha_\beta (\widetilde{h}\widetilde{k})$ can be written when  $\widetilde{h}=\widetilde{k}$ as
\begin{equation}
C^\alpha_\beta (\widetilde{h}\widetilde{k}) = \sum_{i=1}^{N}\sum_{j=1}^{N} w_{ij} (\widetilde{h}\widetilde{k}) E^\alpha_\beta (ij)
\end{equation}
where $w_{ij} (\widetilde{h}\widetilde{k})$ are just real numbers which specify the intensity of the relationship between a node $v_i$ in layer $\widetilde{h}$ and a node $v_j$ in an other layer $\widetilde{k}$.
Then, the fourth-order tensor of the canonical basis of the space $\mathbb{R}^{N \times N \times L \times L}$ is defined as
\label{cab}
\begin{equation}
\xi^{\alpha\widetilde{\gamma}}_{\beta\widetilde{\delta}} (ij\widetilde{h}\widetilde{k}) = E^\alpha_\beta (ij) E^{\widetilde{\gamma}}_{\widetilde{\delta}} (\widetilde{h}\widetilde{k}) =
e^\alpha(i) e_\beta(j) e^{\widetilde{\gamma}}(\widetilde{h}) e_{\widetilde{\delta}}(\widetilde{k})
\label{xi}
\end{equation}
Replacing in Eq.~\ref{muladjtens} the expressions obtained in Eq.~\ref{cab} and Eq.~\ref{xi}, the multilayer adjacency tensor can be written as
\begin{equation}
M^{\alpha\widetilde{\gamma}}_{\beta\widetilde{\delta}} = \sum_{\widetilde{h},\widetilde{k}=1}^{L}\sum_{i,j=1}^{N} w_{ij} (\widetilde{h}\widetilde{k}) \xi^{\alpha\widetilde{\gamma}}_{\beta\widetilde{\delta}} (ij\widetilde{h}\widetilde{k})
\end{equation}

In some cases, it is possible to aggregate multiple networks constructing a single-layer network. This aggregation can be useful in different kind of studies such as those on temporal or social networks. To change a multilayer network into a weighted single-layer network, the corresponding tensor is multiplied by the 1-tensor $U_{\alpha\widetilde{\gamma}}^{\beta\widetilde{\delta}}$. The obtained \textit{projected single-layer network} $P^\alpha_\beta$~\cite{DeDomenico2013} is
\begin{equation}
P^\alpha_\beta = M^{\alpha\widetilde{\gamma}}_{\beta\widetilde{\delta}} U_{\widetilde{\gamma}}^{\widetilde{\delta}} =
\sum_{\widetilde{h}=1}^{L}\sum_{\widetilde{k}=1}^{L} C^\alpha_\beta (\widetilde{h}\widetilde{k})
\end{equation}

A structure similar to the projected single-layer network is the \textit{aggregate} or \textit{overlay single-layer network}~\cite{DeDomenico2013}. It is obtained ignoring the interlayer edges and summing for each node the edges over all layers in the multilayer network. The multilayer adjacency tensor is used to define the aggregate network contracting the indices which correspond to the layer components as 
\begin{equation}
O^\alpha_\beta = M^{\alpha\widetilde{\gamma}}_{\beta\widetilde{\gamma}} =
\sum_{\widetilde{r}=1}^{L} W^\alpha_\beta (\widetilde{r})
\end{equation}
In the case of an aggregate network, the degree computation is the same of a single-layer network and it is computed like in Eq.~\ref{degreesingle}.

On the contrary, the \textit{multidegree centrality vector} $K^\alpha$~\cite{DeDomenico2013} is defined using 1-tensors of the appropriate order as in the case of single-layer networks:
\begin{equation}
K^\alpha = \Big[M^{\alpha\widetilde{\gamma}}_{\beta\widetilde{\delta}} U_{\widetilde{\gamma}}^{\widetilde{\delta}} \Big] u^\beta=
\Big[P^\alpha_\beta \Big] u^\beta =
\Bigg[\sum_{\widetilde{h}=1}^{L}\sum_{\widetilde{k}=1}^{L} C^\alpha_\beta (\widetilde{h}\widetilde{k}) \Bigg] u^\beta =
\sum_{\widetilde{h}=1}^{L}\sum_{\widetilde{k}=1}^{L} k^\alpha(\widetilde{h}\widetilde{k})
\label{multidegree}
\end{equation}
where $k^\alpha(\widetilde{h}\widetilde{k})$ is the degree vector defined in Eq.~\ref{degreesingle} computed on the interlayer adjacency tensor $C^\alpha_\beta (\widetilde{h}\widetilde{k})$.

\section{Dataset}
\label{sec:dataset}

Our dataset (available on Zenodo~\cite{Zenodo2020}) derives from an anti-mafia operation called ``Montagna'' concluded in 2007 by the Public Prosecutor’s Office of Messina (Sicily) and a specialized anti-mafia police unit of the Italian Carabinieri called R.O.S. (Special Operations Group). The investigation focused on two Mafia families known as ``Mistretta'' and ``Batanesi''.
From 2003 to 2007, these families infiltrated several economic activities including the public works in the north-eastern part of Sicily, through a cartel of entrepreneurs close to the Mafia.
The ``Mistretta'' family had also a role of mediator between other families in Palermo and Catania and other criminal organizations around Messina, such as the ``Barcellona'' and the ``Caltagirone'' families. 
\hfill\break\indent 
Our main data source is the pre-trial detention order issued on March 14, 2007 by the judge for preliminary investigations of the Court of Messina. The Court ordered the pre-trial detention for 38 individuals writing a document of more than two hundred pages with an lot of details about the suspects' crimes, activities, meetings, and calls.
\hfill\break\indent 
From the analysis of this document, we initially built two graphs: Meetings and Phone Calls, in which nodes were uniquely associated with suspected criminals and edges specified meetings and phone calls respectively among individuals \cite{Ficara2020, Calderoni2020, Cavallaro2020, Cavallaro2021}. The Meetings graph had $101$ nodes and $256$ edges. The Phone Calls graph had $100$ nodes and $124$ edges. There were $47$ individuals who jointly belonged to both graphs.

\begin{figure}[ht]
\centering
\includegraphics[scale=0.55]{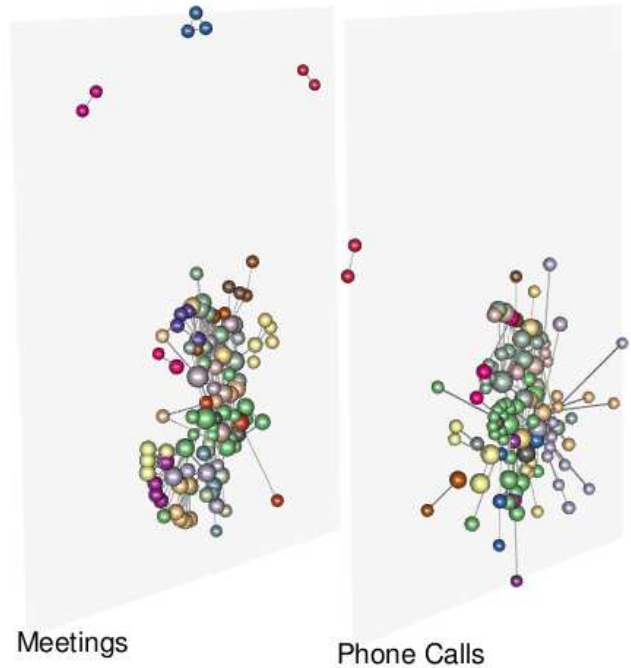}
\caption{\textbf{Multilayer network.} The layered visualization is obtained using Muxviz. The color of nodes is given by their community assignment (\textit{i.e.} how actors are clustered together) and the size by their degree.} 
\label{multilayer}
\end{figure}

Starting from the Meetings and Phone Calls simple graphs, we created an undirected and weighted multilayer network with two layers called Meetings and Phone Calls, $154$ nodes and $439$ edges (see Fig.~\ref{multilayer}). The links within layer Meetings refer to the meetings among members of the criminal network while edges in the layer Phone Calls represent phone communications among distinct phone numbers they use. The weight encodes the number of meetings or phone calls.
According to the definition of the different kind of multilayer networks in Subsect.~\ref{subsec:multi}, we can identify our network as an edge-colored multilayer (\textit{i.e.} a network with multiple types of edges) and more precisely a multiplex network which does not require all nodes to exist on every layer. Each layer have to share at least one node with some other layer in the network to be multiplex. In our case the two layers share $47$ nodes. Moreover, interlayer edges are only those between nodes and their counterparts in another layers and no cost is associated to them.

\section{Methodology and Results}
\label{sec:results}
\begin{figure}[ht]
\centering
\includegraphics[scale=0.5]{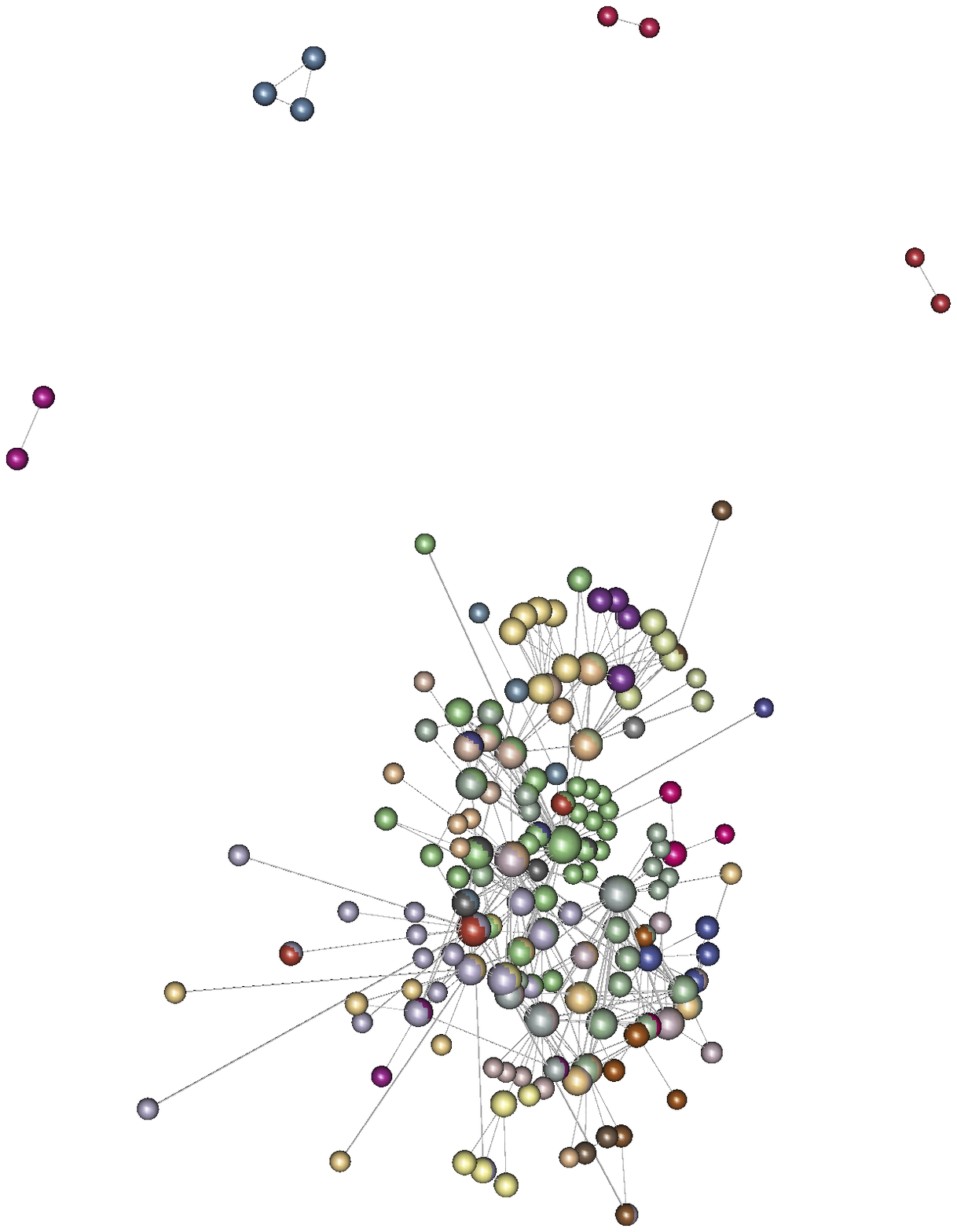}
\caption{\textbf{Aggregate network.} The edge-colored multigraph visualization is obtained using Muxviz. The color of nodes is given by their community assignment (\textit{i.e.} how actors are clustered together) and the size by their degree.} 
\label{aggregate}
\end{figure}

In this paper, we perform an analysis of the nodes' importance in the multiplex network described in Sect~\ref{sec:dataset}. The concept of centrality as ``importance'' is debated and strongly depends on the context. Here we used a simple descriptor which is the degree. This measure quantifies the number of different interaction of each node.
The computation of the nodes' degree in a multilayer network can be done using three different approaches:
\begin{description}
\labitem{Approach 1}{itm:ap1} The two layers of the multilayer network are merged to obtain a single-layer network (\textit{i.e.} the aggregate network shown in Fig.~\ref{aggregate}). This process, often called flattening, is performed creating a new network with one node for every actor and an edge between two nodes if the corresponding actors are connected in any of the layers. Once the aggregate network is obtained, traditional degree (see Eq.~\ref{degreesingle} in Sect.~\ref{sec:background}) can be computed. 
\labitem{Approach 2}{itm:ap2} The traditional degree (see Eq.~\ref{degreesingle} in Sect.~\ref{sec:background}) can be applied to each layer separately. Then, the results are compared.
\labitem{Approach 3}{itm:ap3} Multiple layers are considered at the same time, but without treating them as being ontologically different. Measures based on this approach explicitly consider the difference between interlayer and intralayer edges and also make numerical distinctions between different layers (\textit{e.g.} through weights), but at the end they typically produce single numerical values merging the contributions of the different types of edges \cite{DeDomenico2013, DeDomenico2015} (see Eq.~\ref{multidegree}).
\end{description}

Table~\ref{tab1} gives a summary of the 20 top nodes ranked by their degree in the Aggregate network (\textit{i.e.} according to~\ref{itm:ap1}), in the single layers Phone Calls and Meetings (\textit{i.e.} according to~\ref{itm:ap2}) and in the Multilayer network (\textit{i.e.} according to~\ref{itm:ap3}). The nodes' importance given by their degree is compared with the real roles these nodes have in the Sicilian Mafia families observed during the ``Montagna'' operation. These roles have been reconstructed by us while reading court documents of the ``Montagna'' operation and they are also available on Zenodo~\cite{Zenodo2020}. 

\begin{table}[ht]
\caption{The 20 top ranked nodes in the Multilayer, Aggregate, Phone Calls and Meetings networks compared with their roles in the ``Montagna'' Operation.}
\begin{tabular}{|p{1.1cm}|p{5.5cm}|p{1.8cm}|p{1.8cm}|p{2.1cm}|p{1.6cm}|}
\toprule
\multicolumn{2}{|c|}{\textbf{Node}} 
& \multicolumn{4}{c|}{\textbf{Degree}} \\ 
\midrule
\textbf{Name} & \textbf{Role} & \textbf{Multilayer} & \textbf{Aggregate} & \textbf{Phone Calls} & \textbf{Meetings} \\
\midrule
\midrule
18 & Caporegime Mistretta Family & 51 & 41 & 25 & 24 \\
47 & Deputy Caporegime Batanesi Family & 42 & 29 & 21 & 19\\
27 & Caporegime Batanesi Family & 29 & 21 & 11 & 16 \\
68 & Caporegime Batanesi Family & 27 & 19 &  10 & 15 \\
29 & Enterpreneur & 24 & 16 & 9 & 13 \\
61 & Caporegime Mistretta Family & 23 & 19 & 17 & 4 \\
45 & Associate Batanesi Family & 20 & 14 & 6 & 12 \\
12 & Associate Mistretta Family & 19 & 16 & 1 & 16 \\
11 & Mafia activity coordinator in Messina & 18 & 15 & 4 & 12 \\
22 & Pharmacist & 18 & 15 & 2 & 14 \\
51 & Associate Batanesi Family & 17 & 11 & 4 & 11 \\
25 & Caporegime Mistretta Family & 16 & 13 & 1 & 13 \\
43 & Messaggero & 16 & 11 & 5 & 9 \\
48 & Associate Batanesi Family & 15 & 12 & 1 & 12 \\
19 & External partnership & 14 & 11 & 3 & 9 \\
36 & Aiding and abetting of a fugitive & 14 & 11 & 4 & 8 \\
75 & Associate Mistretta Family & 14 & 12 & 8 & 4 \\
89 & Associate Batanesi Family & 14 & 12 & Absent node & 12 \\
54 & Enterpreneur & 13 & 7 & 5 & 6 \\
5 & Sighted with nodes 11 and 12 & 12 & 10 & Absent node & 10 \\
\bottomrule
\end{tabular} 
\label{tab1}
\end{table}

\begin{figure}[ht]
\centering
\includegraphics[scale=0.6]{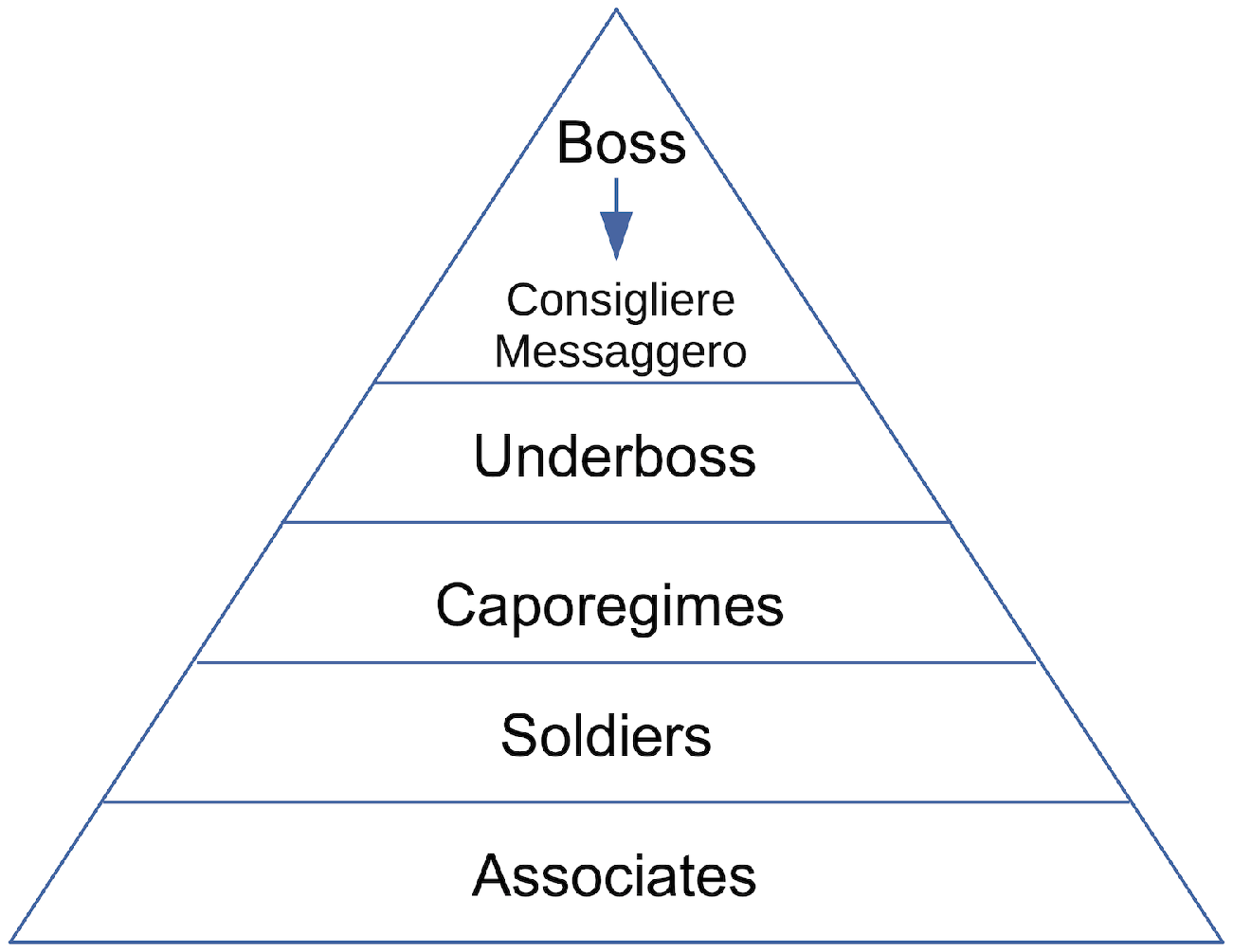}
\caption{The structure of a Mafia Family.} 
\label{roles}
\end{figure}

In particular, a Mafia Family has a typical structure which is shown in Fig.~\ref{roles}. On top of the pyramid hierarchical chart is the \textit{Boss} who makes all the major decisions, controls the Mafia members and resolves any disputes. Usually the real boss keeps a low-profile and keeps his real identity hidden. Just below the boss is the \textit{Underboss} who is the second in command. He can resolve disputes without involving the boss himself and replaces the boss if he is old or in danger of going to jail. In-between the boss and underboss is a role of the \textit{Consigliere} who is an advisor to the boss and makes impartial decisions based upon fairness and for the good of the Mafia. Also in-between there is the  \textit{Messaggero} who is a messenger who functions as liaison between criminal families. He can reduce the need for sit-downs, or meetings, of the mob hierarchy, and thus limits the public exposure of the bosses.
Below the underboss is the \textit{Caporegime} (also called \textit{Captain} or \textit{Capo}) who manages his own crew within the criminal family in a designated geographical location. A Capo's career relies heavily on how much money they can bring into the family. How many capos there are in a given family simply depends on how big that family is. Then, there are \textit{soldiers} who report to their Caporegime. They are street level mobsters who essentially are no more than your average type criminals. Many soldiers can be assigned to one Capo. The final part of a family comes in the shape of \textit{associates}, who are not actual members of the Mafia, but they work with Mafia soldiers and caporegimes on various criminal enterprises. An associate is simply someone who works with the mob, including anyone from a burglar or drug dealer to a pharmacist, entrepreneur, lawyer, investment banker, police officer or politician. 

\begin{figure}[ht]
\centering
\includegraphics[scale=0.8]{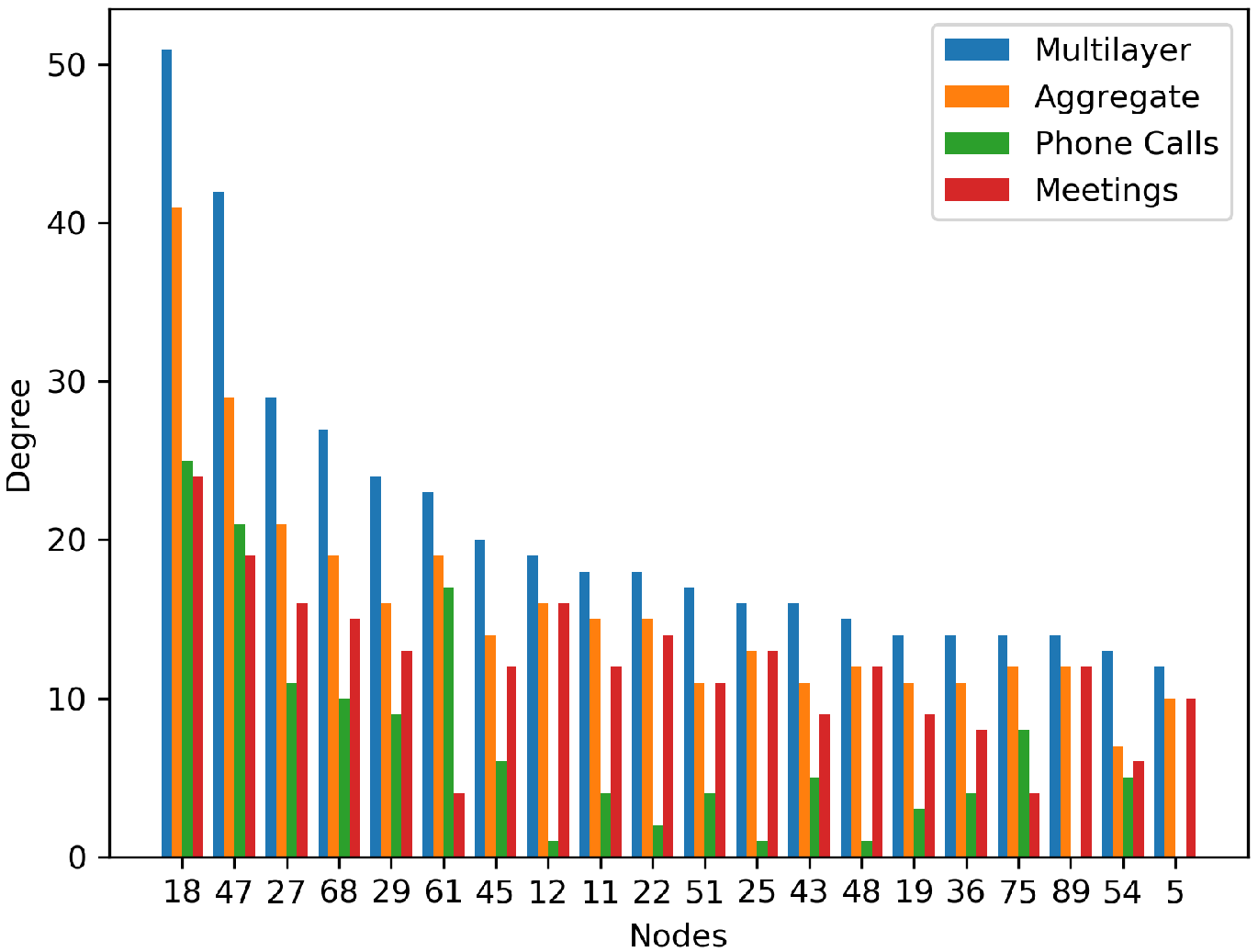}
\caption{The 20 top ranked nodes by degree in Multilayer (Blue), Aggregate (Orange), Phone Calls (Green) and Meetings (Red) networks.} 
\label{MAGPGM}
\end{figure}

We performed the analysis of nodes' degree for each layer separately, the aggregate and the multilayer networks using Muxviz~\cite{MuxViz2014} and Python. The multilayer framework allows to quantify the importance of a node across all the layers. 
The top 20 nodes ranked by their degree are shown in Fig.~\ref{MAGPGM}. The results for each layer separately (\textit{i.e.} \textit{Meetings} or \textit{Phone Calls}), shown in the stacked histogram, reveal that the most important actors per layer are nodes $18$ and $47$.
The result from the aggregate network, obtained by summing up all interactions across the whole network while neglecting the layered structure, also identify nodes $18$ and $47$ as the most central actors.
The same result is obtained for the multilayer network, \textit{i.e.} considering the layered structure.
These two nodes are effectively important because they are respectively Caporegime of the Mistretta family and deputy Caporegime of the Batanesi family.
Using the multilayer framework it was possible to identify two key Caporegimes of the Mistretta family (\textit{i.e.} Nodes $61$ and $25$). Node $61$ was one the twenty most important nodes in the Phone Calls layer but not in the Meetings layer.  Node $25$ was one the twenty most important nodes in the Meetings layer but not in the Phone Calls layer. So, the importance of these nodes doesn't emerge from the analysis of the single layers but only from the analysis of the Aggregate network and even more of the Multilayer one. We can also identify the Messaggero (\textit{i.e.} Node $43$) who didn't seem so important from the analysis of the single layers. Then, we can identify some key associates such us pharmacist or entrepreneurs needed in synthetic drug synthesis processes or to facilitate the award of public contracts to companies close to criminal organizations. The identification of these figures can be very useful to define attack strategies to disrupt criminal networks \cite{Duijn2014, Villani2019, Cavallaro2020}.

\section{Conclusions}
In this paper, we used a real criminal dataset obtained by parsing a two hundred pages pre-trial detection order by the Court of Messina during an anti-mafia operation called ``Montagna'' concluded in 2007. Starting from this dataset we initially built two social networks, one for meetings and one for phone calls among suspects. Since some suspects meet and also call each other, it was natural to identify meetings and phone calls with the layer of a multiplex network. First thing we did was to import the data as an undirected and weighted network where the weight encoded the number of meetings or phone calls.

We wanted to perform an analysis of suspects' importance for each layer separately, for the aggregate network and the multilayer one which allowed to quantify the importance across the whole series of layers. The concept of centrality as ``importance'' is debated and strongly depends on the context. Here we used a simple descriptor which is the degree. This measures just quantifies the number of different meetings or phone calls of each suspect.

We showed the top 20 characters ranked by their degree in each layer, in the aggregate network and in the multilayer network comparing the resulting importance with the role they had in the Sicilian Mafia families which were the protagonists of the ``Montagna'' operation.

As future works, we want to apply to our network other centrality descriptors such as multiplexity which quantifies the fraction of layers where a node appears and PageRank which was introduced by Google's founders and ranks nodes by assuming that more important ones are likely to interact with other important nodes. Then we want to compute the multiplex participation coefficient~\cite{Battiston2014} which quantifies the participation of a node to the different communities of a network.

A further analysis of our dataset revealed the possibility to build a new multilayer network with three layers, one for meetings, one for phone calls and a third layer for the crimes committed by the suspects.

%
%
%
%

\end{document}